\documentclass{article} 
\usepackage{iclr2026_conference,times}

\usepackage{amsmath,amsfonts,bm}

\def\eqref#1{equation~\ref{#1}}

\def\1{\bm{1}}

\DeclareMathAlphabet{\mathsfit}{\encodingdefault}{\sfdefault}{m}{sl}
\SetMathAlphabet{\mathsfit}{bold}{\encodingdefault}{\sfdefault}{bx}{n}

\usepackage{hyperref}
\usepackage{url}
\usepackage{mathpazo}
\usepackage{times}

\usepackage[T1]{fontenc}
\usepackage{amsmath,amssymb,amsfonts}
\usepackage{mathrsfs}
\usepackage{mathabx}
\usepackage{amsbsy}
\usepackage{graphicx}
\usepackage{graphics}
\usepackage{epstopdf}
\usepackage{algorithm}
\usepackage{algorithmic}
\usepackage[algo2e]{algorithm2e}
\usepackage{subfigure}
\usepackage{cite}
\usepackage{multirow}

\usepackage{caption}
\usepackage{tikz,pgfplots}
\usepackage{ctable}
\usepackage{setspace}
\usepackage{mathtools}

\DeclareMathAlphabet{\mathbfsl}{OT1}{ppl}{b}{it} 

\newcommand{\bI}{\mathbfsl{I}}

\newcommand{\bN}{\mathbfsl{N}}

\newcommand{\bX}{\mathbfsl{X}}

\newcommand{\ba}{\mathbfsl{a}}

\newcommand{\bx}{\mathbfsl{x}}

\newcommand*{\rom}[1]{\expandafter\romannumeral #1}

\makeatletter
\newcommand{\AlignFootnote}[1]{%
	\ifmeasuring@
	\else
	\iffirstchoice@
	\footnote{#1}%
	\fi
	\fi}
\makeatother

\newcommand{\be}[1]{\begin{equation}\label{#1}}
\newcommand{\ee}{\end{equation}} 

\newcommand{\script}[1]{{\mathscr #1}}

\newcommand{\Cref}[1]{Co\-ro\-lla\-ry\,\ref{#1}}

\newcommand{\deff}{\mbox{$\stackrel{\rm def}{=}$}}

\title{Learning to Ask: Decision Transformers \\ for Adaptive Quantitative Group Testing}

\author{Mahdi Soleymani, 
	\\
	Halıcıo\u{g}lu Data Science Institute, \\
	University of California San Diego.\\
	\texttt{msoleymani@ucsd.edu} \\
	\And
	Tara Javidi \\
	Halıcıo\u{g}lu Data Science Institute, \\
	University of California San Diego.\\
	\texttt{tjavidi@ucsd.edu} \\
}

\iclrfinalcopy 
\begin{document}

\maketitle

\begin{abstract}
We consider the problem of quantitative group testing (QGT), where the goal is to recover a sparse binary vector from aggregate subset-sum queries: each query selects a subset of indices and returns the sum of those entries. 
 Information-theoretic results suggest that adaptivity  \emph{could} yield up to a twofold reduction in the total number of required queries, yet no algorithm has surpassed the non-adaptive bound, leaving its practical benefit an open question. 
 In this paper, we reduce the QGT problem to an integer-vector recovery task whose dimension scales with the sparsity of the original problem rather than its full ambient size. We then formulate this reduced recovery task as an offline reinforcement learning problem and employ Decision Transformers to solve it adaptively. By combining these two steps, we obtain an effective end-to-end method for solving the QGT problem. Our experiments show that, for the first time in the literature, our adaptive algorithm reduces the average number of queries below the well-known non-adaptive information-theoretic bound, demonstrating that adaptivity can indeed reduce the number of queries.
\end{abstract}

\section{Introduction}

Quantitative Group Testing  (QGT) is the problem of detecting $k$ defective items within a collection of $n$ items through a series of tests conducted on $m$ distinct pools.  Each test returns an integer indicating how many defective items are present in the pooled subset. QGT has been widely applied in areas such as rare variant detection in genome sequencing \citep{cao2014quantitative}, network traffic monitoring \citep{wang2015group}, resource allocation in random access communication \citep{de2021optimal}, and structured signal recovery \citep{matsumoto2023improved, mazumdar2021support}.

Strategies for solving the QGT problem are typically categorized as \emph{adaptive} or \emph{non-adaptive}, depending on whether the design of future tests can incorporate information from earlier test outcomes.  In the non-adaptive approach, all tests must be predetermined and can be executed in parallel that enables parallelization  at the cost of having more queries. In contrast, in the adaptive approach, one can observe the results of previous tests and utilize this information to design subsequent tests . Adaptive strategies can reduce the total number of tests by refining queries based on previously observed outcomes, but they require dynamically updating the test design after each round of results. From an information-theoretic perspective, there exists a factor of $2$ gap between the lower bounds on the minimum number of queries required in adaptive and non-adaptive settings. This implies that adaptively designing tests \emph{could} reduce the total number of queries by up to half compared to the non-adaptive approach. We consider the adaptive case in this paper. 

Decision Transformers (DTs) introduced by \citep{chen2021decision}, are a class of sequence modeling approaches that frame decision-making tasks as a conditional sequence prediction problem. Decision Transformers were originally proposed for reinforcement learning that model trajectories of the form consisting of triplets of \emph{return-to-go}, \emph{state}, and \emph{action} using a transformer architecture which enables learning patterns from offline dataset without explicitly requiring value functions or policy optimization. This method leverages the ability of transformers to model long-range dependencies that effectively captures complex temporal patterns and generalizes across a wide range of tasks. This formulation is particularly well-suited for problems with structured sequential data and variable-length decision processes, making it a promising approach for complex combinatorial tasks like quantitative group testing. 
 Follow-up works have extended DTs to offline settings with broad trajectory modeling \citep{janner2021offline}, multi-task and multi-game learning \citep{lee2022multi}, and few-shot policy generalization \citep{xu2022prompting}. Notably, recent efforts such as  \citep{brohan2022rt1} scaled transformer-based policies to real-world robot control at massive scale. These successes inspire our investigation of DTs in combinatorial settings, where strategic long-term planning and reward-guided decision-making play a central role.

A straightforward approach to solving the QGT problem is to frame it as a reinforcement learning (RL) task and train a Decision Transformer (DT) accordingly. Although this method can be effective for small problem sizes, that is, small $n$, it becomes intractable as $n$ grows large. This is because the action space grows exponentially with $n$; specifically, the agent must choose from $2^n$ possible actions at each step. Such exponential growth poses a significant scalability challenge, a common obstacle in many combinatorial optimization problems. As $n$ increases, both the size of the model and the amount of data required to train it also grow exponentially, often rendering machine learning approaches impractical. This motivates us to transfer the the original problem, which lives in an ambient $n$-dimensional space, into a significantly smaller $k$-dimensional one. This dimensionality reduction enables the use of modern machine learning techniques on combinatorial problems without suffering from scalability issues.
We then leverage Decision Transformers trained on millions of solved instances of the reduced problem and use them as a core component to solve the original QGT problem through adaptive queries. Our extensive experiments demonstrate that the performance gap between adaptive and non-adaptive strategies can be effectively closed. Notably, for $k = 2$, our method achieves the optimal number of queries for the first time in the literature. More broadly, our approach is the first to surpass the known non-adaptive bounds using a data-driven adaptive strategy.

\section{Related Work}

The QGT problem has been extensively studied under both probabilistic and combinatorial models. In the combinatorial setting, the number of defectives $k$ is fixed and known, while in the probabilistic model, each item is defective independently with some probability $p$. Our work aligns with the combinatorial model. One category of solutions to QGT  rely on non-adaptive testing schemes, which design all tests in advance. The information-theoretic lower bound on the number of tests required in a non-adaptive scheme for  the QGT problem in this regime is \citep{djackov}
\be{lb}
m_0\deff \frac{2k}{\log k}\log \frac{n}{k}
.\ee 
Several works have leveraged connections to compressed sensing to develop recovery guarantees under linear programming and convex optimization \citet{donoho2006compressed, candes2006robust}, but these typically require $\Theta(k \log n)$ tests. More specialized QGT algorithms reduce this test count by a factor of $O(\log k)$, but leave a logarithmic gap to the information-theoretic lower bound \citet{karimi2019sparse, feige2020quantitative, gebhard2022parallel, tan2023approximate}. To narrow this gap, \citet{hahn2022near,soleymani2024non} propose non-adaptive schemes achieving \emph{almost} optimal number of tests with polynomial decoding complexity. These contributions highlight the trade-off between computational efficiency and test optimality. 

On the adaptive side, a simple counting argument yields a lower bound that is half of the one in (\ref{lb}), e.g., see \citep{bshouty2009optimal}. However, no existing algorithm has been able to effectively leverage adaptivity to surpass the non-adaptive bound. The best known result is due to \citet{bshouty2009optimal}, who proposed an algorithm that achieves a number of tests within a factor of 2 of the optimal value. In this work, we aim to tackle this gap by leveraging Decision Transformers in combination with a  reduction approach. Specifically, we aim to design an adaptive querying process that achieves a query count between $m_0$ and the adaptive lower bound $\frac{m_0}{2}$, which is the lower bound for adaptive sterategies.



\section{Problem Setting}
\begin{figure}[t]
\centering\small
\includegraphics[width=0.6\columnwidth]{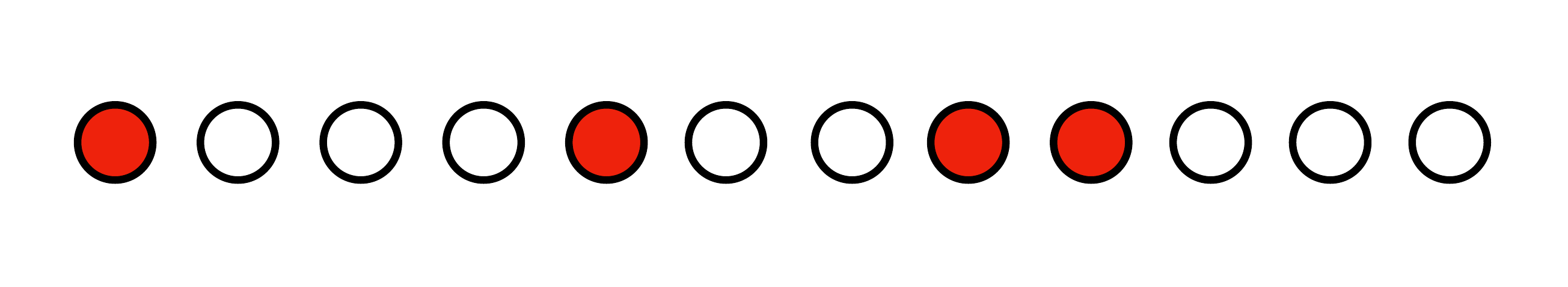} 
\caption{An instance of the QGT problem, where a subset of items includes a few defective ones (shown in red). The corresponding incident vector is 
$\bx= [1,0,0,0,1,0,0,1,1,0,0,0]$, indicating the positions of the defective items. }
\label{ill}
\end{figure}

Pooled data recovery problem \citet{dorfman1943detection, aldridge2019group, mazumdar2016nonadaptive, cheraghchi2011graph, emad2014poisson} refers to the task of identifying unknown categories or values from aggregated queries over over selected pools/subsets of items. Queries return only a pooled result rather than individual values. In this paper, we focus on the binary case, also known as the QGT problem, in which a set of $n$ items is given out of which $k$ items are defective. However, our methods extend to more general settings with suitable adaptations.  The \emph{incident} vector corresponding to these items is a binary vector $\bx \in \{0,1\}^n$ such that $\bx_i=1$ if the item is defective and $\bx_i=0$, otherwise. Let $\script{B}_{n,k}$ denote the set of all binary vectors of size $n$ with  $k$ non-zero elements. In this paper, we consider the combinatorial model for the pooled data recovery problem where $\bx$ is sampled uniformly at random from $\script{B}_{n,k}$. 
The goal is to identify all defective items, that is, recover $\bx$, by performing measurements on as few subsets of the items as possible. The \emph{outcome} of a test is the number of defective items belonging to the underlying subset.  This problem is also known as the \emph{coin weighing problem} and \emph{quantitative group testing problem} in the literature. Figure\,\ref{ill} provides a graphical illustration of this abstract problem where red circles represent defective items, i.e., the coordinates of $\bx$ with value $1$.

In the following section, we introduce a binary splitting algorithm that reformulates the original problem as an integer vector recovery task of size $k$, rather than 
$n$. This reduction allows us to cast the problem within a reinforcement learning framework, where the action space scales with $k$ which is a constant under the sparse setting thereby decoupling the algorithm's complexity from the size of the pooled data recovery task. We further leverage the Decision Transformer model to solve this problem adaptively using offline datasets. Since the model does not influence the data generation process, the training procedure remains highly efficient regardless of the data generation method or complexity.

\section{Binary Splitting: Reformulating the Problem in Integer Space}\label{split}

A key challenge in utilizing modern machine learning method to combinatorial problems is the rapid growth in complexity as the problem size increases. This often renders standard ML approaches impractical since the required model capacity, training data, and computational resources scale exponentially with the size of the problem. 
 To address this, we introduce a reduction strategy for the QGT problem: instead of solving for a binary vector of length $n$, we reduce the problem to  identifying a sparse integer-valued vector of length $k$. As $k$ is often a constant, this reduction dramatically shrinks the effective problem size, making it far more amenable to learning-based techniques.
This key insight allows us to leverage machine learning methods to reduce the number of required measurements beyond what is possible with existing combinatorial techniques. In the following section, we describe our approach and demonstrate how offline reinforcement learning, specifically Decision Transformers, can be effectively applied to this reduced problem.

Our reduction method is the following. Initially, we partition the $n$ items into $k$ disjoint groups and use a query budget of $k$ to measure the number of defective items in each group. Next, we split each group into two halves. We then form a new vector of length $k$ by aggregating the \emph{first half} of each group. The number of defectives in each of these half-groups defines an \emph{integer-valued vector}, where each coordinate corresponds to the defective count in one half-group. Note that the size of this new vector is $k$, and we already know an upper bound for each coordinate; these are the results of our initial measurements on the $k$ full groups before splitting. Thus, identifying the number of defectives in the first-half splits reduces to recovering an integer-valued vector of length $k$ with known upper bounds $u_1, u_2, \ldots, u_k$, where each $u_i$ corresponds to the total number of defectives in the $i$-th original group.

\begin{figure*}[t]
\centering
\includegraphics[width=0.7\textwidth]{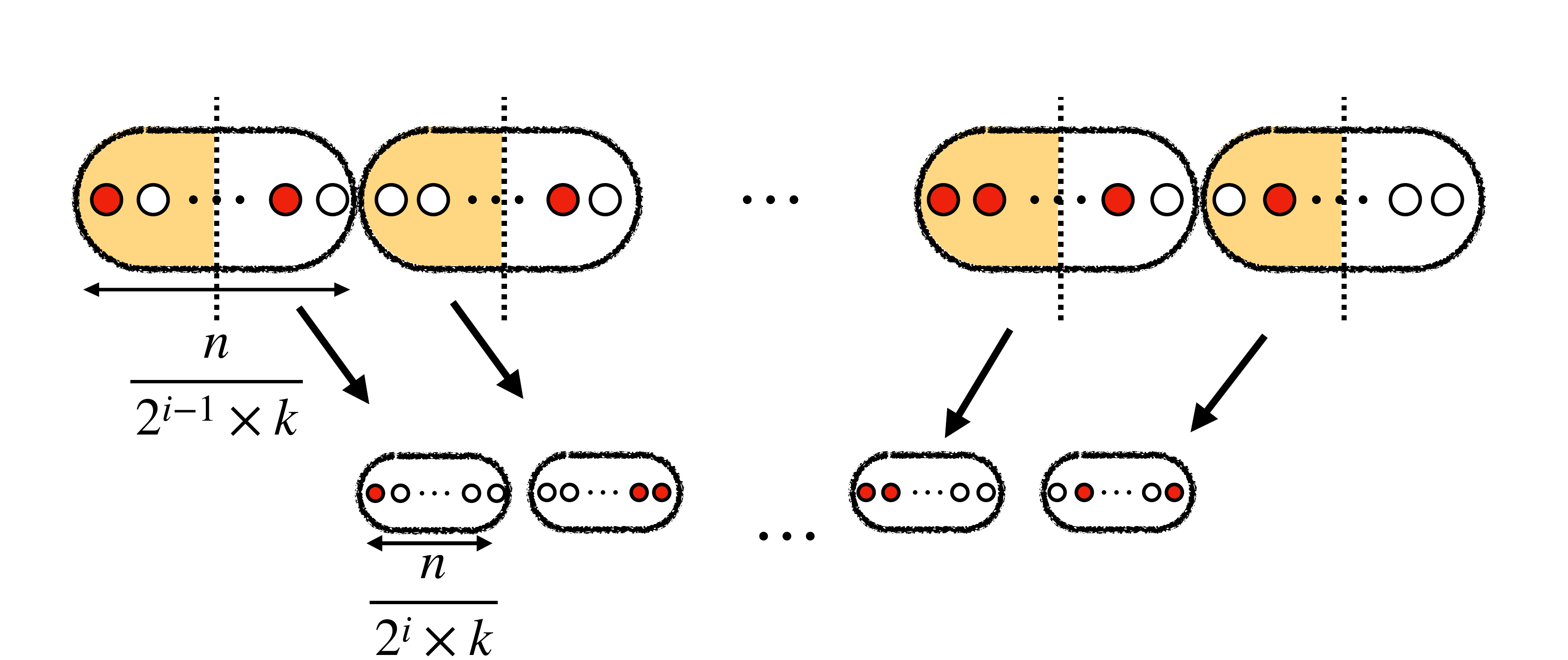} 
\caption{Illustration of the recursive splitting procedure at stage $i$. Each group is divided into two halves, and the left-half splits (highlighted) are treated as a new subproblem. Once the number of defectives in the left halves is recovered using a solver (described in the next section), the defective counts in the right halves can be obtained by subtraction.
}
\label{divide}
\end{figure*}

Assuming access to a solver (or model) that, given a set of upper bounds and a sequence of queries, can efficiently recover the corresponding integer-valued vector, we can determine the number of defectives in the \emph{second-half} splits by subtracting the recovered values from the original group totals. After this procedure, we obtain at most $k$ groups containing non-zero numbers of defectives. In effect, this step reduces the size of each group by half while preserving the total number of active groups, i.e., groups with defectives, at most $k$. The splitting process is demonstrated in Figure\,\ref{divide} for stage $i$.

This splitting procedure can be applied recursively: in each stage, every group is divided into two halves, and the algorithm is applied to the left halves. Once the size of each group reaches $1$, all defective items are localized, and the identification process is complete. Given that the initial group size is $\frac{n}{k}$, the total number of splitting stages required is $\log_2\left(\frac{n}{k}\right)$.

Let $M_i$ denote the number of queries used by the solver at the $i$-th splitting stage. The total query complexity of the entire process is 
$
\sum_{i=1}^{\log_2\left(\frac{n}{k}\right)} M_i + k,
$
where the additive $k$ corresponds to the initial set of measurements used to determine the group-wise upper bounds. 
Assuming that the query cost $M_i$ at each stage follows a similar distribution, we denote the expected value as $\mathbb{E}[M_i] = m$. This assumption holds under the condition that the coordinates of the target vector are randomly shuffled at the beginning, which we ensure during preprocessing. The average total number of queries is therefore:

$$
\log_2\left(\frac{n}{k}\right) \cdot m + k.
$$

In the next section, we focus on training a model that adaptively selects queries based on the entire history of query-result pairs. The objective is to minimize $m$, thereby reducing the total number of measurements required to recover the original binary vector.


\section{Decision Transformers for Integer-Valued Vector Identification}\label{int_vec}



In this section, we develop an adaptive algorithm to solve the reduced problem resulting from the recursive splitting procedure described in Section\ref{split}. Specifically, the task is to identify an \emph{integer-valued vector} of length $k$, given an associated upper bound vector of the same size that specifies the maximum  value for each coordinate. We frame this task as an \emph{offline reinforcement learning} problem, where the agent sequentially selects queries with the objective of identifying the target vector using as few queries as possible.

To address this, we leverage the Decision Transformer framework introduced by \citet{chen2021decision}. In the original formulation, the Decision Transformer is trained on sequences of $\text{(return-to-go, state, action)}$ triplets. These sequences are input into a transformer model, which is trained to predict the next action at each timestep, conditioned on the trajectory history. The model is trained to minimize the difference between the predicted and actual actions, effectively learning to imitate high-return decisions condition on past trajectories.
In our context, we adapt this framework by replacing the state-action pairs with ( return-to-go, query result, query) triplets, enabling the model to learn an effective query policy from offline data.

\noindent\textbf{Trajectory representation.}  Let $(r_t, s_t, \mathbf{a}_t)$ denote the reward, state, and action at timestep $t$. We consider a binary reward model, where the agent receives a reward of $-1$ at each step until the target vector is uniquely identified, at which point the reward becomes $0$. That is, the agent is penalized for each additional query made before solving the problem.
In the Decision Transformer framework, instead of using the immediate reward directly in the input sequence, the model operates on the return-to-go (RTG), defined as the cumulative sum of all future rewards from timestep $t$ onward:
$
\hat{R}_t \triangleq \sum_{t'=t}^{T} r_{t'}.
$
The state $s_t$ at each timestep corresponds to the response (i.e., the sum result) of the previous query. At the beginning of the sequence, when no queries have yet been made, the state is initialized to a fixed constant value, which we set to $k$, the total number of defective items.
The action $\mathbf{a}_t$ is a binary vector of length $k$, representing the query made at timestep $t$. A value of $1$ in position $i$ indicates that the $i$-th item is included in the query pool, while a value of $0$ indicates it is not.

\begin{figure}[t]
\centering
\includegraphics[width=0.7\columnwidth]{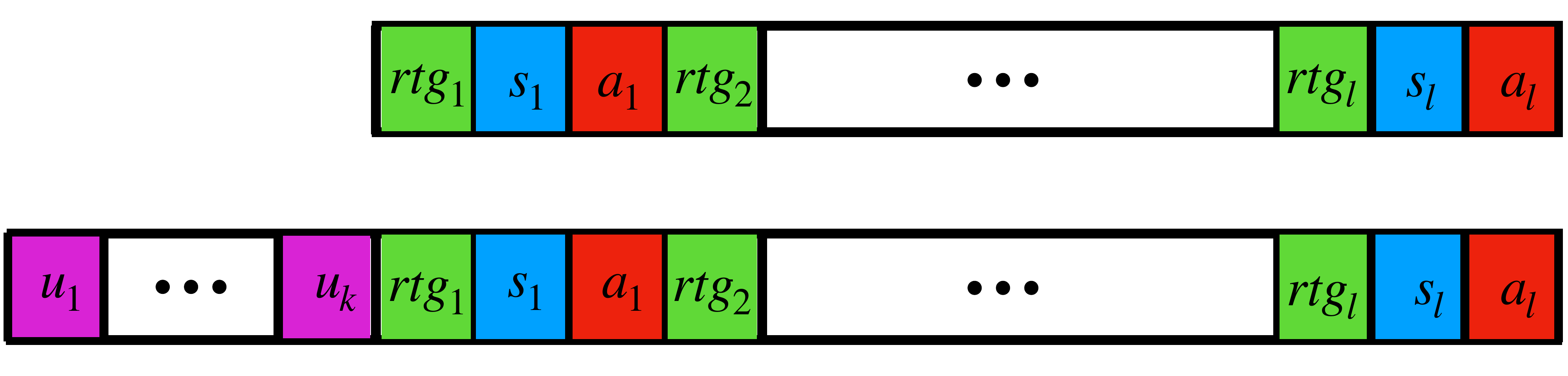} 
\caption{
Illustration of the input sequence structure for the GPT model. The top sequence follows the structure used in \citet{chen2021decision}, adapted to our problem setting. In the bottom sequence, we additionally prepend the known upper bounds to the input, providing the model with information about the specific instance being solved.
}
\label{seq}
\end{figure}

\noindent\textbf{Architecture.}
Transformers, introduced by  \citet{vaswani2017attention}, are powerful models for processing sequential data. They rely on a self-attention mechanism that allows each element in a sequence to weigh the relevance of all other elements it its outputs. This enables the model to capture long-range dependencies and contextual relationships effectively. Similar to \citet{chen2021decision}, we use a GPT-style transformer \citet{radford2018improving}, which adds a causal attention mask so that each output depends only on previous inputs. This design is  suitable for our setting, where the model must generate the next query based solely on the history of previous queries, responses, and return-to-go values.

We explore two input sequence designs for our Decision Transformer model. In the first design, we follow the structure proposed by \citet{chen2021decision}, where a sequence of length $l$ consisting of $( return\text{-}to\text{-}go, \ state, \ action )$ triplets is fed into the model. This results in a total of $3l$ input tokens, with $l$ treated as a tunable design parameter.

In the second design, we incorporate additional  information by including the vector of known upper bounds. Specifically, we prepend the upper bound values (represented as a sequence of integers of length $k$) to the beginning of the input sequence. This results in a total of $3l + k$ tokens. The motivation behind this design is to provide the model with information about the structure of the specific instance it is solving, allowing it to tailor its query strategy more effectively to the underlying problem. The structure of both input formats is illustrated in Figure\,\ref{seq}.

Each token in the input sequence is first passed through a modality-specific linear layer to produce learned embeddings. These embeddings are then projected into a common embedding space, followed by layer normalization. The resulting sequence is processed by a GPT-style transformer \citet{radford2018improving}, which autoregressively predicts the next action token.

\begin{figure}[t]
\centering
\includegraphics[width=0.7\columnwidth]{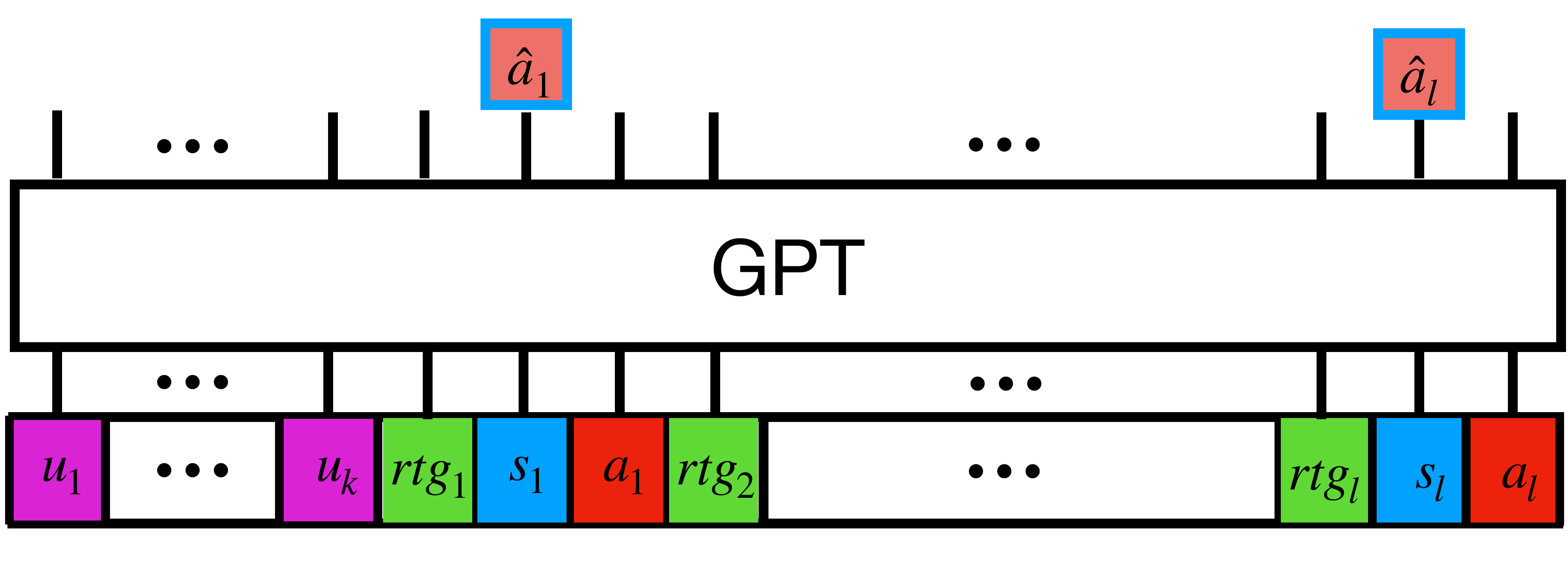} 
\caption{
Illustration of how sequences are fed into the GPT model and how the outputs are used. We focus only on the model outputs corresponding to state tokens, ignoring the rest. At each state token, the model is trained to predict the corresponding action $\hat{a}\_i$ as accurately as possible. This reflects the model’s ability to suggest the next query based on the full history of prior return-to-go, query results, and actions. Each action is a binary vector of length $k$, indicating which items are included in the next query.}
\label{GPT}
\end{figure}

\noindent\textbf{Training.}
We feed input sequences of length $3l$ or $3l + k$, depending on the scenario, drawn from the dataset. The prediction head associated with each input token corresponding to a state $s_t$ is trained to predict the corresponding action $a_t$ using cross-entropy loss. See Figure\,\ref{GPT}. The losses across all timesteps in the sequence are averaged to compute the final training loss. We experimented with alternative loss functions, including mean squared error (MSE), Huber loss, and angular loss, but did not observe any improvement in the model’s overall performance.

\noindent\textbf{Evaluation.} During evaluation, we must specify the initial return-to-go, which in our setting corresponds to the negative of the number of steps expected to uniquely identify the target vector. Since this number is unknown in advance, we treat it as a tunable parameter. To optimize performance, we perform a sweep over values ranging from $-1$ to $-l$, where $l$ is the maximum sequence length (i.e., the number of query-response-return triplets allowed). The return-to-go is incremented by $1$ at each step. If it reaches $0$ before the vector has been uniquely identified, we continue to feed $-1$ as the return-to-go in subsequent steps until the problem is solved. This ensures the model continues to issue queries even if the initial return-to-go estimate was insufficient, reflecting the uncertainty in how many steps are required to solve a specific instance.


\section{Experiments}



\begin{table*}[t]
\centering
\caption{Average number of queries vs. query strategies. The first row shows the performance of a random agent. The second row corresponds to a DT agent trained on the trajectory of a random agent. The third and fourth rows correspond to DT agents trained on expert datasets that maximize covariance and conditional entropy at each stage, respectively. The baseline corresponds to the results in \citet{bshouty2009optimal}, which are the best known to date. The lower bound represents the fundamental information-theoretic limit for the minimum number of queries needed.}
\begin{tabular}{l|ccccccc}
\toprule
\textbf{Method} 
& \textbf{k=2} 
& \textbf{k=3} 
& \textbf{k=4} 
& \textbf{k=5} 
& \textbf{k=6} 
& \textbf{k=7} 
& \textbf{k=8} \\
\midrule
\textbf{Random} & 2.42 & 3.15 & 3.78 & 4.22 & 4.74 & 5.03 & 5.49 \\
\textbf{DT-Random} & 1.78 & 2.58 & 3.27 & 3.83 & 4.29 & 4.83 & 5.11 \\
\textbf{DT-Covariance Guided} & 1.28 & 1.96 & 2.30 & 3.23 & 3.76 & 4.34 & 4.97 \\
\textbf{DT-Entropy Guided} & \textbf{1.26} & \textbf{1.79} & \textbf{2.29} & \textbf{2.95} & \textbf{3.47} & \textbf{3.94} & \textbf{4.30} \\
\textbf{Baseline} & \textbf{2.52} & \textbf{3.00} & \textbf{3.45} & \textbf{3.87} & \textbf{4.28} & \textbf{4.67} & \textbf{5.05} \\
\textbf{Lower Bound} & 1.26 & 1.50 & 1.72 & 1.93 & 2.14 & 2.33 & 2.52 \\
\bottomrule
\end{tabular}
\label{tab:query_performance}
\end{table*}

In this section, we report the performance of the Decision Transformer trained on several datasets for the integer vector identification problem described earlier. 
We begin by generating random trajectory sequences for training. At each step, a binary query vector is sampled uniformly at random, independent of prior queries or responses. We then determine whether the target vector is uniquely identifiable given the set of queries so far.

To check identifiability, we use the Gurobi solver via the \emph{gurobipy} Python library \citet{gurobi}. Specifically, the solver is used only to verify whether the current set of linear equations admits a unique solution. Note that due to our reduction method discussed earlier, this verification operates in a reduced space of dimension $k$ (the sparsity of the original problem), rather than the original ambient dimension $n$. This allows us to use the solver efficiently as a decoding oracle, since the computational cost depends only on $k$, which is fixed, rather than on $n$, which is often large.

In our first experiment, we train the Decision Transformer on a dataset of completely random trajectories. Each sequence follows the structure illustrated in  the top of Figure\,\ref{seq}, where only return-to-go, state, and action triplets are included which mirrors the original setup proposed in \citet{chen2021decision}. The average number of queries required to solve the problem, is demonstrated in the row specified by DT-Random column in Table\,\ref{tab:query_performance}.  The results partially align with the findings in \citet{chen2021decision}, indicating that the model can  indeed extract meaningful patterns from entirely random trajectories. Specifically, it performs better than purely random querying. However, the improvement is incremental and there is still a significant gap to the lowerbound for adaptive scenarios.  This is likely due to the increased complexity of our problem compared to, for example, finding the shortest pass on random graphs considered in \citet{chen2021decision} where solutions are often unique. In contrast, our problem admits many different optimal query sequences, including permutations of the same query set. Since the Decision Transformer attempts to learn an optimal sequence, the presence of multiple equivalent solutions increases the ambiguity and makes the learning task significantly more challenging.
DFCĆ1QW

For the remainder of this paper, we adopt the input sequence structure that incorporates upper bounds, as it provides the model with additional problem-specific context. This choice consistently improves performance, as confirmed by our experiments across all datasets. In our third experiment, we train the model on a more structured \emph{expert} dataset, specifically one generated using a so-called \emph{covariance-based} strategy. In this approach, queries are selected to maximize the covariance with respect to the underlying hidden vector. Let $\bX$ represent the random hidden integer vector and $\Sigma_i$ denote the empirical covariance matrix estimated after $i$ queries. This empirical covariance matrix can be estimated as $\bN^T\bN$, where $\bN$ is a matrix whose rows consist of all feasible solutions that are consistent with the query results. Then, the next query is chosen by solving the following optimization problem:

$$
\ba_{i+1} \deff \arg\max_{\bI \in \{0,1\}^k} \bI^T \Sigma_i \bI,
$$

a quadratic integer program that can be solved using efficient solvers such as \textit{Gurobi} via the \texttt{gurobipy} library in Python. We then train the model on these covariance-guided trajectories. As shown in the DT-Covariance Guided row in Table\ref{tab:query_performance}, the model outperforms the one trained on random trajectories, benefiting from the more informative training data and learning more meaningful patterns as a result.

Finally, we generate an \emph{entropy-guided} dataset, where at each step the next query is selected to maximize the conditional entropy of the hidden vector $\bX$, given all previous queries and their results. To estimate the conditional entropy, we heuristically sweep over all feasible solutions for a candidate query, count the frequency of each possible query result, and construct the empirical distribution. The entropy is then computed based on this distribution. Although this method is significantly more time-consuming than the random and covariance-guided approaches, it produces highly informative trajectories. When trained on this entropy-guided dataset, our model achieves the best performance across all settings. As shown in  the DT-Entropy Guided row in Table\ref{tab:query_performance}, this setup yields the strongest results observed in our experiments. In particular, for $k=2$, the model learns to achieve the lower bound, indicating that it has effectively captured all available information, as it is not possible to reduce the number of queries beyond this point.

All synthetic datasets used in our experiments consist of $10$ million offline-generated trajectory samples. Due to the sequential nature of trajectory construction, this data generation process cannot be parallelized and may become time-consuming. However, this delay scales no faster than linearly in 
$k$. More significantly, for covariance- and entropy-guided strategies, additional computations such as estimating empirical covariance matrices or conditional entropies, introduce substantial overhead. These operations result in latency that grows exponentially with 
$k$, making such methods increasingly impractical for inference on larger problem sizes.  In contrast, training a GPT-based model to \emph{imitate} these strategies offers significant advantages in terms of latency. At inference time, our model avoids these expensive computations entirely and benefits from GPU-based parallelism, making our method  scalable with respect to $k$. 
We compare the latency of each method in Table\,\ref{tab:latency}. The results show that our method significantly reduces inference time compared to expert agents, such as those optimizing covariance and conditional entropy, thereby making it a more practical and scalable solution. All experiments were conducted using CPU only on a Macbook Pro. We did not exploit our method’s ability to run in parallel on GPU, which can significantly reduce the latency of the DT-based approach when deployed on a GPU, depending on the specific hardware. This can, in principle, widen the gap between the DT agent and the covariance- and entropy-maximizing agents to an arbitrary extent.
\section{Conclusion}

The implication of combining the results from the previous section with the reduction method discussed in Section \ref{split} for the QGT problem is that it can be solved using the value shown in Table \ref{tab:query_performance} times the $\log_2\frac{n}{k}$ factor, which comes from the binary splitting step. In particular, for $k = 2$ our method achieves, for the first time in the literature, the optimal value matching the information-theoretic lower bound. Moreover, for all other values of $k$, our method surpasses the non-adaptive lower bound which is the best baseline known to date. This experimentally demonstrates, for the first time, that adaptivity can indeed be leveraged to exceed the fundamental limits of non-adaptive algorithms for the QGT problem. These results motivate further investigation into adaptive strategies, ideally aiming to close the gap to the adaptive information-theoretic lower bound.


\begin{table}
\centering
\caption{Average latency in milliseconds vs. $k$ for different query strategies. The first row shows the latency of our DT agent using CPU only. The second and third rows show the latencies for agents that maximize covariance and conditional entropy, respectively. }
\begin{tabular}{l|ccccccc}
\toprule
\textbf{Agent} 
& \textbf{$k$=2} 
& \textbf{$k$=3} 
& \textbf{$k$=4} 
& \textbf{$k$=5} 
& \textbf{$k$=6} 
& \textbf{$k$=7} 
& \textbf{$k$=8} \\
\midrule
\textbf{DT-Agent} & \textbf{40} & \textbf{53} & \textbf{58} & \textbf{75} & \textbf{107} & \textbf{127} & \textbf{173} \\
\textbf{Cov-Agent} & 280 & 310 & 350 & 400 & 460 & 520 & 590 \\
\textbf{Ent-Agent} & 330 & 370 & 410 & 450 & 500 & 570 & 640 \\
\bottomrule
\end{tabular}
\label{tab:latency_transposed}
\end{table}

\pagebreak 
\bibliography{iclr2026_conference}
\bibliographystyle{iclr2026_conference}


\end{document}